\documentclass[aps,prl,twocolumn,superscriptaddress,showpacs]{revtex4}
\usepackage{graphicx}
\usepackage{latexsym}
\usepackage{amssymb}
\usepackage{amsmath}
\usepackage{amsfonts}
\usepackage{bm}
\usepackage{multirow}
\usepackage{color}
\usepackage{comment}
\newcommand{\ii}{\mathrm{i}}

\newcommand{\eqnref}[1]{Eq.\,\eqref{#1}}

\newcommand{\beq}{\begin{equation}}
\newcommand{\eeq}{\end{equation}}
\newcommand{\beqn}{\begin{eqnarray}}
\newcommand{\eeqn}{\end{eqnarray}}

\DeclareMathAlphabet{\mathbbold}{U}{bbold}{m}{n}

\makeatletter
\newcommand\xleftrightarrow[2][]{%
\ext@arrow 9999{\longleftrightarrowfill@}{#1}{#2}}
\newcommand\longleftrightarrowfill@{%
\arrowfill@\leftarrow\relbar\rightarrow} \makeatother

\begin{document}

\title{Emergent Symmetry and Tricritical Points near the deconfined \\ Quantum Critical Point}

\author{Chao-Ming Jian}
\affiliation{Kavli Institute of Theoretical Physics, Santa
Barbara, CA 93106, USA} \affiliation{ Station Q, Microsoft
Research, Santa Barbara, California 93106-6105, USA}
\author{Alex Rasmussen}
\affiliation{Department of Physics, University of California,
Santa Barbara, CA 93106, USA}
\author{Yi-Zhuang You}
\affiliation{Department of Physics, Harvard University, Cambridge,
MA 02138, USA}
\author{Cenke Xu}
\affiliation{Department of Physics, University of California,
Santa Barbara, CA 93106, USA}

\date{\today}
\begin{abstract}

Recent proposal of the duality between the $N=2$ noncompact
QED$_3$ and the easy-plane noncompact CP$^1$ (NCCP$^1$) model
suggests that the deconfined quantum critical point (dQCP) between
the easy-plane antiferromagnet and the VBS order on the square
lattice may have an emergent O(4) symmetry, due to the
self-duality of the $N=2$ noncompact QED$_3$. Recent numerical
progresses suggest that this easy-plane dQCP does exist and it has
an emergent O(4) symmetry. But for the O(4) symmetry to really
emerge at the dQCP, certain O(4) symmetry breaking perturbations
need to be irrelevant at the putative O(4) fixed point. It is more
convenient to study these symmetry breaking perturbations in the
$N=2$ noncompact QED$_3$. We demonstrate that a natural large-$N$
generalization and a controlled $1/N$ expansion supports the
stability of the O(4) fixed point against the symmetry breaking
perturbations. We also develop the theory for two tricritical
points close to the easy-plane dQCP. One tricritical point is
between the dQCP and a {\it self-dual} $Z_2$ topological order;
the other is the tricritical point that connects the continuous
dQCP and a first order N\'{e}el-VBS transition, motivated by
recent numerical results.

\end{abstract}

\maketitle

\begin{figure}[tbp]
\begin{center}
\includegraphics[width=200pt]{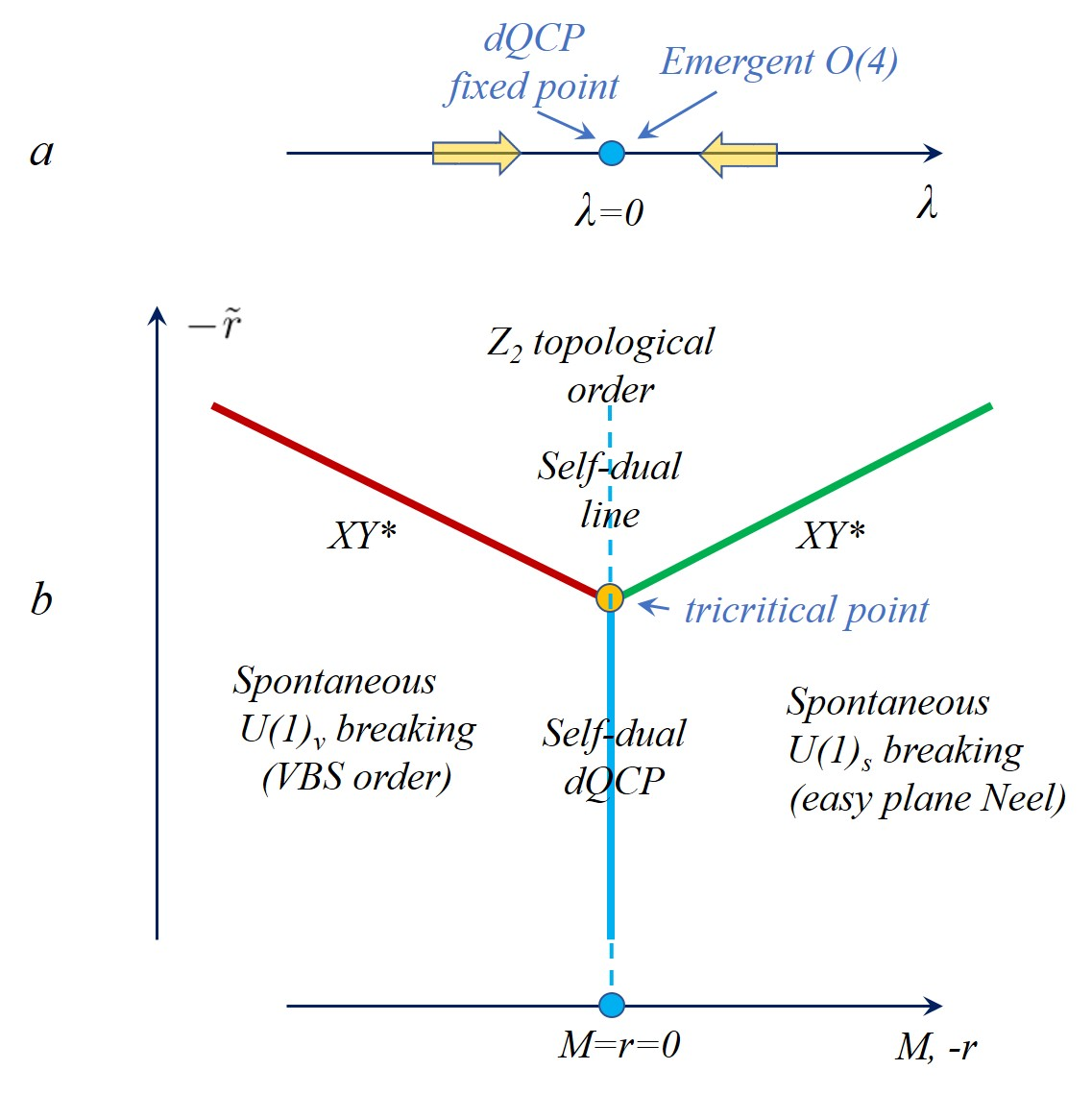}
\caption{($a$) Our RG equation Eq.~\ref{RG} suggests that the
perturbation $\lambda$ in Eq.~\ref{4f2} that breaks the O(4)
symmetry down to $[\mathrm{O(2)}_s \times \mathrm{O(2)}_v] \times
Z_2^d$ at the self-dual dQCP is irrelevant, which supports the
emergence of O(4) symmetry at the infrared limit of the dQCP, and
is consistent with recent numerical results; ($b$) The sketched
phase diagram of Eq.~\ref{yukawa2}, plus the tuning parameter $r$,
or $M$ from Eq.~\ref{duality}. Especially, across a tricritical
point, the system enters a {\it self-dual $Z_2$ topological order
where the self-dual symmetry $Z_2^d$ exchanges the $e$ and $m$
anyons.} } \label{PD}
\end{center}
\end{figure}

Recent progress of $(2+1)d$ conformal field theories (CFT) has led
us to expect that different Lagrangians at their quantum critical
points may correspond to the same CFT, $e.g.$ a property called
``duality". Within these proposed dualities, one is of great
importance to condensed matter theory, which is the duality
between the $N=2$ noncompact QED$_3$ and the easy-plane NCCP$^1$
model at the critical point~\cite{karchtong,potterdual,SO5}. These
two field theories can be written as~\footnote{We will take the
Euclidean space-time, and choose the following convention for the
$\gamma_\mu$ matrices throughout: $(\gamma_0, \gamma_1, \gamma_2)
= (\sigma^2, \sigma^3, \sigma^1)$.}
\begin{subequations}
\label{duality}
\begin{align}
& \mathcal{L}_\text{QED} =
\sum_{j=1}^2 \bar{\psi}_j \gamma\cdot (\partial - \ii a) \psi_j + m \bar{\psi}_j \psi_j + M \bar{\psi}\sigma^3 \psi \label{duality1} \\
& \mathcal{L}_\text{CP$^1$} = \sum_{j=1}^2 |(\partial - \ii
b)z_j|^2 + g |z_j|^4 + r |z_j|^2 + h z^\dagger \sigma^3 z
\label{duality2}
\end{align}
\end{subequations}
where $\psi_j$ and $z_j$ are two-component Dirac fermions (with an
extra flavor index $j$) and complex boson fields coupled to
non-compact U(1) gauge fields, $a_\mu$ and $b_\mu$, respectively.
The duality maps the variables $(m,M)$ to $(h,r)$.

When realized in terms of lattice quantum many-body systems, the
tuning parameter $r$ of Eq.~\ref{duality2} drives a phase
transition between the easy-plane N\'{e}el order and a valence
bond solid (VBS) order, and it is called the deconfined quantum
critical point (dQCP)~\cite{deconfine1,deconfine2}. Despite the
earlier numerics which suggest a first order easy-plane
N\'{e}el-to-VBS transition~\cite{planar1,planar2,planar3}, most
recently a modified lattice model was found which did show a
continuous easy-plane dQCP~\cite{epjq} (there were more numerical
evidences for the continuous dQCP with isotropic SO(3) spin
symmetry~\cite{dqcp1,dqcp2,dqcp3,dqcp4,dqcp5,dqcp6,dqcp7,dqcp8,dqcp9,dqcp10,dqcp11}).

On the other hand, theoretically the tuning parameter $m$ in
Eq.~\ref{duality1} drives the phase transition between the bosonic
symmetry protected topological phase and the trivial
phase~\cite{Tarun_PRB2013,lulee}, and it was shown numerically
that such transition is also second order, as long as the system
has high enough symmetries~\cite{kevinQSH,mengQSH2}.

Before the more recent proposal of duality Eq.~\ref{duality}, it
was first shown in Ref.~\cite{ashvinlesik} that Eq.~\ref{duality2}
is self-dual at its critical point $r = 0$ and $ h = 0$. This
self-duality can be derived by performing the particle-vortex
duality for each flavor of
$z_j$~\cite{peskindual,halperindual,leedual}, followed by
integrating out the gauge field $b_\mu$. Thus at the critical
point $r = 0$, $h = 0$, the field theory has an explicit symmetry
$[\mathrm{O}(2)_s \times \mathrm{O}(2)_v] \times Z^d_2$.
The $\mathrm{O}(2)_s = U(1)_s \rtimes Z^s_2$ symmetry is the
inplane spin rotation symmetry that acts on the CP$^1$ field
$(z_1, z_2)^t$ as \beqn U(1)_s: z \rightarrow e^{ i
\frac{\theta}{2} \sigma^3 } z, \ \ \ Z^s_{2}: z \rightarrow
\sigma^1 z. \eeqn The $\mathrm{O}(2)_v = U(1)_v \rtimes Z^v_2$
symmetry corresponds to the conservation and particle-hole
symmetry of the gauge flux of $b_\mu$: \beqn U(1)_v: \mathcal{M}_b
\rightarrow e^{i \theta} \mathcal{M}_b, \ \ \ Z^v_2 :
\mathcal{M}_b \rightarrow \mathcal{M}_b^\dagger, \ z \rightarrow i
\sigma^2 z^\dagger,\eeqn where $\mathcal{M}_b$ is the monopole
operator ($2\pi-$gauge flux annihilation operator) of the gauge
field $b_\mu$. The last $Z^d_2$ corresponds to the self-duality
transformation which interchanges the two O(2) symmetries, and it
precludes the $r$ term in Eq.~\ref{duality2} if $Z^d_2$ is imposed
as an actual symmetry. The $h$ term is excluded by the $Z^s_2$
symmetry, which is the improper rotation subgroup of O(2)$_s$.

Eq.~\ref{duality1} was shown to be also self-dual in
Ref.~\cite{xudual,mross,seiberg2}, by performing the fermionic
version of the particle-vortex
duality~\cite{son2015,wangsenthil1,wangsenthil2,maxashvin,seiberg1}
for each flavor of Dirac fermion $\psi_1$ and $\psi_2$
individually and integrating out $a_\mu$. This self-duality
suggests that the infrared symmetry of both Eq.~\ref{duality1} and
Eq.~\ref{duality2} at $r = h = m = M = 0$ (assuming these two
theories are both conformal field theories at this point, as was
suggested by recent
numerics~\cite{kevinQSH,mengQSH2,Karthik2016,epjq}) could be as
large as O(4)$\sim \mathrm{SO}(4) \times Z_2$, where
$\mathrm{SO}(4)$ corresponds to the product of the SU(2) flavor
symmetries of both sides of the self-duality of
Eq.~\ref{duality1}, and the $Z_2$ improper rotation is the
self-dual transformation of the $N=2$ noncompact QED$_3$, and it
is equivalent to either $Z_2^s$ or $Z_2^v$ (which rotate to each
other under $\mathrm{SO}(4)$). In the dQCP theory
Eq.~\ref{duality2}, the corresponding O($4$) order parameter is
\beqn \boldsymbol{N} = \left( z^\dagger \sigma^x z, z^\dagger
\sigma^y z, \mathrm{Re}[\mathcal{M}_b], \mathrm{Im}[\mathcal{M}_b]
\right). \label{vector} \eeqn However, the O(4) emergent symmetry
is not immediately obvious in \eqnref{duality2}. If the O(4)
symmetry indeed emerges at the easy-plane NCCP$^1$ critical point,
then the O(4) invariant fixed point must be stable against
symmetry breaking perturbations that break O(4) down to its
microscopic symmetry $[\mathrm{O}(2)_s \times \mathrm{O}(2)_v ]
\times Z_2^d$.

The symmetry-breaking perturbation can be most conveniently
analyzed in the $N=2$ noncompact QED formalism, and it corresponds
to one four fermion interaction term \beqn \lambda
\left(\bar{\psi} \sigma^3 \psi \right)^2 = \frac{\lambda}{2}
\left( \sigma^1_{ij} \bar{\psi}_{i \alpha} \epsilon_{\alpha\beta}
\bar{\psi}_{j\beta}\right)\left( \sigma^1_{ij} \psi_{i \alpha}
\epsilon_{\alpha\beta} \psi_{j\beta}\right) + \cdots
\label{4f1}\eeqn The ellipses are terms that preserve the SU(2)
flavor symmetry of the $N=2$ QED$_3$, such as $(\bar{\psi}\psi)^2$
and $(\bar{\psi}\gamma^\mu\psi)^2$. This term Eq.~\ref{4f1} breaks
the symmetry of the $N = 2 $ noncompact QED$_3$ down to the
desired $[\mathrm{O}(2)_s \times \mathrm{O}(2)_v] \times Z^d_2$
symmetry. The generators of $U(1)_s$ and $U(1)_v$ are two
different linear combinations of the remaining $U(1)_A$ flavor
symmetry of Eq.~\ref{duality1} generated by $\sigma^3$, and the
$U(1)_B$ symmetry that corresponds to the conservation of the flux
of $a_\mu$: \beqn U(1)_s &:& \psi \rightarrow e^{ i
\frac{\theta}{2} \sigma^3 } \psi, \ \ \mathcal{M}_a \rightarrow
e^{ i \frac{\theta}{2} } \mathcal{M}_a , \cr\cr U(1)_v &:& \psi
\rightarrow e^{ - i \frac{\theta}{2} \sigma^3 } \psi, \ \
\mathcal{M}_a \rightarrow e^{i \frac{\theta}{2}} \mathcal{M}_a
\label{f42} \eeqn The $Z_2^s$ and $Z_2^v$ symmetries involve the
self-duality transformation of the $N=2$ QED$_3$, while their
product $Z_2^{s \times v} = Z_2^s \times Z_2^v$ flips the charge
of both $U(1)_s$ and $U(1)_v$, and it acts as $Z_2^{s \times v}:
\psi \rightarrow \psi^\dagger$, $a_\mu \rightarrow - a_\mu$. The
self-dual $Z^d_2$ transformation of Eq.~\ref{duality2} corresponds
to the ``flavor flipping" symmetry $\psi \rightarrow \sigma^1
\psi$. (For more details of how the symmetries act on the $N=2$
noncompact QED$_3$, please refer to Ref.~\cite{SO5})

It appears that another four fermion term $\sum_\mu
\left(\bar{\psi} \sigma^3 \gamma_\mu \psi \right)^2$ is allowed
once we break the symmetry of the $N=2$ QED$_3$ down to
$[\mathrm{O}(2)_s \times \mathrm{O}(2)_v] \times Z_2$. But this
term is not linearly independent from the term in Eq.~\ref{4f1}
and two other $\mathrm{SU}(2)$ symmetric terms when $N=2$. To
analyze whether the symmetry breaking term is relevant or not at
the $N=2$ noncompact QED$_3$ fixed point, we need a controlled
calculation of its scaling dimension. And like many previous
studies of $(2+1)d$ QED$_3$, a large$-N$ generalization and a
$1/N$ expansion is very helpful for this purpose.

In Ref.~\cite{herbut,xusachdev,pufu}, a large$-N$ generalization
of Eq.~\ref{duality1} was taken, and a $1/N$-expansion calculation
of the scaling dimensions of SU($N$) invariant four fermion
interaction perturbations suggest that these SU($N$) invariant
four fermion terms are likely always irrelevant even for small
$N$.
However, in Ref.~\cite{herbut,xusachdev} it was also shown that
once we break the SU($N$) flavor symmetry of the QED$_3$, some
four fermion interaction may become relevant for small enough $N$,
and lead to instability of the QED$_3$. Thus we need to test
whether the symmetry breaking term Eq.~\ref{4f1} causes this
potential concern. But to evaluate this we need a large$-N$
generalization of Eq.~\ref{4f1}. For this purpose, we change the
basis and consider the interaction
$\lambda(\bar{\psi}\sigma^2\psi)^2$, which then has a natural
large$-N$ generalization: \beqn \frac{g}{N} \sum_{i,j}
\left(\bar{\psi}_i \psi_j \right)\left(\bar{\psi}_i \psi_j
\right). \label{4f2} \eeqn This term breaks the global symmetry of
noncompact QED$_3$ with $N$ flavors of Dirac fermions down to
$\mathrm{O}(N) \times \mathrm{O}(2)$, where the O(2) corresponds
to the conservation and particle-hole symmetry of the gauge flux
of $a_\mu$. 
The advantage of this large$-N$ generalization is that, there is
also only one term that breaks the symmetry down to $\mathrm{O}(N)
\times \mathrm{O}(2)$, for arbitrary $N$. Another seemingly
$\mathrm{O}(N) \times \mathrm{O}(2)$ invariant four-fermion term $
\sum_\mu \sum_{i,j} \left(\bar{\psi}_i \gamma_\mu \psi_j
\right)\left(\bar{\psi}_i \gamma_\mu \psi_j \right)$ is a multiple
of Eq.~\ref{4f2} after using the Fierz identify of $\gamma_\mu$.

Unfortunately, the self-duality of the original $N=2$ noncompact
QED$_3$ no longer holds in this large$-N$ generalization. Despite
the disadvantage of losing the self-duality, the same method as
Ref.~\cite{xusachdev} leads to the following RG equation for $g$
at the leading order of the $1/N$ expansion: \beqn \frac{dg}{dl} =
\Big(- 1 - \frac{64}{3 N \pi^2}\Big)g + O(g^2). \label{RG}\eeqn
This means that the first order $1/N$ correction to $g$ makes it
even more irrelevant. This calculation is consistent with the
recent numerical observation that an easy-plane J-Q
model~\cite{epjq}, a model that has a continuous transition
between the easy-plane N\'{e}el and VBS order has the same set of
critical exponents as another model with an exact microscopic
SO(4) symmetry, hence both models are supposed to have an emergent
O(4) symmetry at the critical point, meaning the perturbations
that break the O(4) to $[\mathrm{O}(2)_s \times \mathrm{O}(2)_v]
\times Z^d_2$ is irrelevant.

The four fermion term Eq.~\ref{4f2} is perturbatively irrelevant
at the $N=2$ noncompact QED$_3$ fixed point, but when the
microsopic perturbation that leads to Eq.~\ref{4f2} is strong
enough, it can lead to new physics. For example, when $\lambda$ is
negative, its effect can be captured by the following Lagrangian:
\beqn \mathcal{L}_\text{QED-Yukawa} &=& \sum_{j=1}^2 \bar{\psi}_j
\gamma\cdot (\partial - \ii a) \psi_j + u \bar{\psi}\sigma^3 \psi
\phi \cr\cr &+& (\partial_\mu \phi)^2 + \tilde{r} \phi^2 + g
\phi^4. \label{yukawa1}\eeqn $\phi$ is a real scalar field. When
$\tilde{r} > 0$, $\phi$ is in its disordered phase, and
integrating out $\phi$ will generate a short range four fermion
interaction term Eq.~\ref{4f2}, which as we evaluated above is an
irrelevant perturbation at the $N=2$ noncompact QED$_3$ fixed
point. When $\tilde{r} < 0$, $\phi$ will be ordered, and the
system spontaneously generates an expectation value of $\phi$.
Recalling that the mass term $M \bar{\psi}\sigma^3 \psi$ is the
tuning parameter of the N\'{e}el-VBS phase transition, thus when
$\tilde{r}<0$, the system spontaneously breaks the ``self-dual"
symmetry of the easy-plane NCCP$^1$ model, and the N\'{e}el-VBS
phase transition becomes first order. Thus $\tilde{r} = 0$ is a
tricritical point between the continuous easy-plane deconfined QCP
and a first order N\'{e}el-VBS transition, which is an analogue of
the tricritical Ising fixed point.
This tricritical point between a continuous and discontinuous
easy-plane N\'{e}el-VBS transition was first discussed in
Ref.~\cite{ribhu} in the formalism of NCCP$^1$ field theory, and
our Lagrangian Eq.~\ref{yukawa1} can be viewed as the dual
description of this tricritical point. Recent numerical simulation
of one particular class of easy-plane spin-1/2 model on the square
lattice also suggests the existence of this tricritical
point~\cite{epjq}.

Another tricritical point near the dQCP can be described by the
following QED-Yukawa-Higgs type of Lagrangian: \beqn
\mathcal{L}^\prime_\text{QED-Yukawa} &=& \sum_{j=1}^2 \bar{\psi}_j
\gamma\cdot (\partial - \ii a) \psi_j + u \left( \sigma^1_{ij}
\psi_{i \alpha} \epsilon_{\alpha\beta} \psi_{j\beta}\right) \phi
\cr\cr &+& H.c. + |(\partial - \ii 2a) \phi|^2 + \tilde{r}
|\phi|^2 + g |\phi|^4. \label{yukawa2}\eeqn Now $\phi$ is a
complex scalar field instead of a real scalar. Again, when
$\tilde{r}$ is positive, $\phi$ is disordered, and system is
described by Eq.~\ref{duality1} plus irrelevant short range
four-fermion interaction Eq.~\ref{4f2}; while when $\tilde{r} <
0$, $\psi$ forms a Cooper pair condensate, and the U(1) gauge
field $a_\mu$ is Higgsed and broken down to a $Z_2$ gauge field.

To understand exactly the phase with $\tilde{r}<0$, let us first
analyze its symmetry. The Cooper pair $\left( \sigma^1_{ij}
\psi_{i \alpha} \epsilon_{\alpha\beta} \psi_{j\beta}\right)$
preserves the $U(1)_A$ flavor symmetry of $\psi_j$ generated by
$\sigma^3$, and since the photon is Higgsed and gapped, the
$U(1)_B$ symmetry which corresponds to the conservation of the
gauge flux is also preserved. Since $U(1)_s$ and $U(1)_v$ are
combinations of these two U(1) symmetries, both $U(1)_s$ and
$U(1)_v$ are preserved. The $Z_2^{s\times v} = Z_2^s \times Z_2^v$
symmetry is also obviously preserved even in the condensate,
because in the condensate of $\phi$, the particle-hole
transformation of the expectation value $\langle \phi \rangle$ can
be cancelled by a gauge transformation.

Obviously the condensate of $\phi$ will gap out all the fermions,
and the photon $a_\mu$ acquires a Higgs mass, thus this phase is
fully gapped. The gapped excitations of this phase include a
fermion $\psi$, which carries a $Z_2$ gauge charge, and the
$U(1)_A$ flavor symmetry, which is a combination of $U(1)_s$ and
$U(1)_v$. The $\pi-$flux of $a_\mu$ (the so-called vison) which is
bound with a vortex of $\phi$ is another gapped excitation. The
quantum number of the vison can be extracted by solving the Dirac
equation with a background vortex of $\phi$, and we can see that
there is one complex fermion zero mode at the vortex core. Each
vortex core will carry the $U(1)_B$ quantum number of $\pi-$flux
of $a_\mu$, and $\pm 1/2$ quantum number of the flavor $U(1)_A$
charge carried by the fermion zero mode. Thus these two visons
with filled and unfilled fermion zero modes will carry half charge
under $U(1)_s$ and half charge under $U(1)_v$ respectively.
Because these two types of visons differ by a fermion, they will
have mutual semion statistics caused by the Aharonov-Bohm effect
between the fermion and the $\pi-$flux. For the same reason, the
two types of visons also carry the same topological spins. This is
because their difference in topological spins is the sum of the
topological spin of the extra fermion and the Aharonov-Bohm phase
between the extra fermion and the $\pi-$flux, which cancel each
other.

Now let us label the $\pi-$flux carrying half $U(1)_s$ charge as
the $e$ particle, and label the $\pi-$flux carrying half $U(1)_v$
charge as the $m$ particle. Usually, different topological
excitations have different energies. However, in this case, since
the $Z_2^d$ self-dual symmetry is unbroken, the $e$ and $m$
particles transform into each other under the $Z_2^d$ symmetry,
and hence, are degenerate. A more concrete way of showing this
degeneracy is to understand the previously mentioned complex
fermion zero mode more carefully. When $\phi$ condenses, each of
the Dirac fermions with $\sigma^1 = \pm 1$ eigenvalues, denoted as
$\psi_\pm$ respectively, forms a copy of the Fu-Kane
superconductor~\cite{fukane}, while both are coupled to the same
$Z_2$ gauge field. Therefore, the vortex of $\phi$ will carry two
Majorana zero modes $\gamma_\pm$ from the two copies of the
Fu-Kane superconductors. These two Majorana zero modes form the
previously mentioned complex fermion zero mode. Generically, the
two Majorana zero modes can hybridize and, as a consequence, lift
the complex fermion zero mode. Such hybridization can, for
example, be induced by a finite $M \bar{\psi} \sigma^3 \psi$ term.
However, when the $Z_2^d$ symmetry is preserved, any hybridization
of the two Majonrana zero modes are prohibited because they carry
different charges under $\sigma^1$ or equivalently under $Z_2^d$:
\beqn Z_2^d: \gamma_+ \to \gamma_+, \ \ \gamma_- \to -\gamma_-.
\eeqn Therefore, the degeneracy between the $e$ and $m$ particle
is ensured by $Z_2^d$. Also, the occupation number $\ii \gamma_+
\gamma_-$ of the complex fermion zero modes (constructed from the
two Majorana zero modes), which distinguishes the $e$ and $m$
particles, changes under the action of $Z_2^d$. Therefore, we can
conclude that the $Z_2^d$ symmetry exchanges the $e$ and $m$
particles.

Since the $Z_2^d$ self-dual symmetry is unbroken, $e$ and $m$
particles are transformed into each other under the $Z_2^d$
symmetry. Also, since the $Z_2^{s \times v}$ is unbroken, each $e$
and $m$ are a doublet, because $Z_2^{s \times v}$ perform a
particle-hole transformation on both $U(1)_s$ and $U(1)_v$, and
$e$ and $m$ can both carry $+1/2$ or $-1/2$ of their respective
$U(1)$ symmetry. Or in other words, $e$ and $m$ carry projective
representation of O(2)$_s$ and O(2)$_v$ respectively.


We can also understand the condensate of $\phi$ in another way. In
a single slab geometry of a $3d$ TI, when we consider a Fu-Kane
superconductor on its top surface and a time-reversal breaking
bottom surface, this slab can be identified with a $p_x+ip_y$ or
$p_x-ip_y$ superconductors depending on the time-reversal breaking
pattern on the bottom surface.
Now, in the condensate phase of $\phi$ in theory Eq.
\ref{yukawa2}, we are effectively dealing with two copies of the
(gauged) Fu-Kane superconductors. We can equivalently think of
this phase as a phase hosted by two copies of the TI slabs. On the
top surfaces of both slabs we consider the Fu-Kane superconductors
as we did before. But on the bottom surfaces of two TI slabs, we
demand them have opposite time-reversal breaking patterns. When
the fermions are coupled to the same gauge field $a_\mu$, the
Chern-Simons terms of the gauge field $a_\mu$ coming from the two
bottom surfaces cancel each other, leaving the total topological
order of the two slabs exactly that of two gauged Fu-Kane
superconductors on the top surfaces of the two TI slabs.

This picture is very helpful for the understanding of the
topological order of the $\phi$ condensate.
Suppose we coupled the fermions in the two TI slabs to two
independent $Z_2$ gauge fields, we would get a $\rm{Ising} \times
\overline{\rm{Ising}}$ topological order. An Ising topological
order have anyon $1$, $\sigma$, $f$, which are vacuum, nonabelian
anyon, and a fermion respectively, while the anyon content,
labelled by $1$, $\bar{\sigma}$, $\bar{f}$ is similar in the
$\overline{\rm{Ising}}$ topological order. To recover the
condensed phase of $\phi$ in Eq.~\ref{yukawa2}, we need to set the
gauge fields in the two TI slabs equal (and identify them with the
gauge field $a_\mu$ in Eq. \ref{yukawa2}), which can be enforced
by condensing the $f \bar{f}$ particle in the $\rm{Ising} \times
\overline{\rm{Ising}}$ topological order. The topological order
induced by the condensate of $f \bar{ f}$ is exactly a $Z_2$
topological order~\cite{bais2009}. In the condensate, the fermions
$f$ and $\bar{f}$ are identified with each other and also with the
fermion excitation $\psi$. The $\sigma \bar{\sigma}$ particle is
also deconfined in the $f \bar{ f}$ condenstate. In fact, it
splits into two Abelian particles,
and these two Abelian anyons are exactly the two types of visons
$e$ and $m$ introduced before. Also, the topological spins of the
two visons are inherited from that of the $\sigma \bar{\sigma}$
particle, which is trivial (bosonic). All other topological
excitations, $\sigma_\pm$ for instance, in the $\mathrm{Ising}
\times \overline{\mathrm{Ising}}$ topological order are confined
and hence will not appear the condensate of $f \bar{f}$.

Now let us summarize our results: when tuning $\tilde{r}$ in
Eq.~\ref{yukawa2} from positive to negative, the system enters a
$Z_2$ topological order, with bosonic and mutual semionic $e$ and
$m$ particles carrying projective representation of O(2)$_s$ and
O(2)$_v$ symmetries respectively. The $Z_2^d$ duality symmetry
interchanges the $e$, $m$ particles.

The $N=2$ noncompact QED$_3$ was proposed as the boundary state of
the $3d$ bosonic symmetry protected topological (SPT)
state~\cite{xudual}. And it has been known that the boundary of
many $3d$ bosonic SPT states could be a $2d$ $Z_2$ topological
order with $e$ and $m$ particles carrying anomalous quantum
numbers~\cite{senthilashvin,xuclass} (the symmetry
$[\mathrm{O}(2)_s \times \mathrm{O}(2)_v ] \times Z_2^d$ of
Eq.~\ref{duality2} is anomalous if viewed as an on-site symmetry,
and it is a subgroup of the SO(5) symmetry which also supports a
$3d$ bosonic SPT state~\cite{SO5}). The $Z_2$ spin liquid for
spin-1/2 systems on the square lattice that preserve the square
lattice symmetry has also been discussed
recently~\cite{subirz2,faz2}, thus it is conceivable that this
$Z_2$ spin liquid is not so far away from the easy-plane dQCP, and
in that system $e$ is the standard bosonic spinon, while $m$ will
carry lattice momentum hence its condensate will lead to the VBS
order~\cite{maz2gauge,subirz2gauge,sondhiz2gauge,xuz2gauge}.

We can also build exactly the same $Z_2$ topological order using
the dual theory of Eq.~\ref{duality1}, by condensing the Cooper
pair of the dual Dirac fermions. Starting with Eq.~\ref{duality2},
a more standard way to enter this $Z_2$ topological order, is by
first breaking the $Z_2^d$ symmetry and spontaneously breaking the
$U(1)_s$ or $U(1)_v$ symmetry, and then condense the double vortex
of the $U(1)$ order parameter to restore the symmetries. This is
equivalent to condensing the singlet pair of $z_j$ in
Eq.~\ref{duality2}, which was discussed in detail in
Ref.~\cite{subirz2,faz2}. The phase transition between the $Z_2$
topological order and the standard spontaneous U(1) symmetry
breaking phase (superfluid) is the so-called $3d$ XY$^\ast$
transition~\cite{balentsz2,senthilz2,isakov,melko1,melko2}. In
this procedure, the final topological order has the $Z_2^s$ and
$Z_2^v$ symmetry (or the self-duality of Eq.~\ref{duality1}), but
eventually we need to adjust the energy gap for $e$ and $m$ to
restore the $Z_2^d$ self-dual symmetry of Eq.~\ref{duality2}.
Eq.~\ref{yukawa2} shows how to connect the easy-plane dQCP to the
$Z_2$ spin liquid, {\it while preserving the self-dual $Z_2^d$
symmetry.} Our results are summarized in the sketched phase
diagram Fig.~\ref{PD}$b$.

While finishing the current paper, the authors became aware of two
independent upcoming works that partially relate to our current
paper (Ref.~\cite{alexupcoming,senthilupcoming}). Cenke Xu and
Alex Rasmussen are supported by the David and Lucile Packard
Foundation and NSF Grant No. DMR-1151208. Chao-Ming Jian is funded
by the Gordon and Betty Moore Foundation's EPiQS Initiative
through Grant GBMF4304.

\bibliography{deconfine}

\end{document}